\documentclass[conference]{IEEEtran}
\IEEEoverridecommandlockouts
\usepackage{cite}
\usepackage{amsmath,amssymb,amsfonts}
\usepackage{algorithmic}
\usepackage{graphicx}
\usepackage{textcomp}
\usepackage{textcomp}
\usepackage{booktabs} 
\usepackage{url}
\usepackage{tikz}
\usetikzlibrary{shapes.geometric, arrows.meta, positioning, shadows, fit, calc}
\usepackage{tcolorbox}
\usepackage{xcolor}
\usepackage{color}
\def\BibTeX{{\rm B\kern-.05em{\sc i\kern-.025em b}\kern-.08em
    T\kern-.1667em\lower.7ex\hbox{E}\kern-.125emX}}
\begin{document}

%\title{Adaptive Cybersecurity Learning for Criminal Justice Students and Professionals Through a Think–Plan–Act AI Chatbot with Pedagogical Scaffolding *\\
\title{CyberJustice Tutor: An Agentic AI Framework for Cybersecurity Learning via Think–Plan–Act Reasoning and Pedagogical Scaffolding *\\
%{\footnotesize \textsuperscript{1}}
\thanks{This research was supported, in part, by National Science
Foundation (NSF) Grants 2334196 and 2334197.}
}

\author{\IEEEauthorblockN{Baiqiang Wang}

\IEEEauthorblockA{School of Engineering and Technology \\
University of Washington Tacoma\\
Tacoma, US \\
wbq@uw.edu}
~\\
\and
\IEEEauthorblockN{Yan Bai}
\IEEEauthorblockA{School of Engineering and Technology \\
University of Washington Tacoma\\
Tacoma, US \\
yanb@uw.edu}
%*Corresponding author
~\\
\and
\IEEEauthorblockN{Juan Li}
\IEEEauthorblockA{Department of Computer Science\\
North Dakota State University\\
Fargo, US \\
j.li@ndsu.edu}
}

\maketitle

\begin{abstract}
The integration of Large Language Models (LLMs) into cybersecurity education for criminal justice professionals is currently hindered by the "statelessness" of reactive chatbots and the risk of hallucinations in high-stakes legal contexts. To address these limitations, we propose the CyberJustice Tutor, an educational dialogue system powered by an Agentic AI framework. Unlike reactive chatbots, our system employs a "Think–Plan–Act" cognitive cycle, enabling autonomous goal decomposition, longitudinal planning, and dynamic context maintenance. We integrate a Pedagogical Scaffolding Layer grounded in Vygotsky’s Zone of Proximal Development (ZPD), which dynamically adapts instructional support based on the learner's real-time progress. Furthermore, an Adaptive Retrieval Augmented Generation (RAG) core anchors the agent's reasoning in verified curriculum materials to ensure legal and technical accuracy. A comprehensive user study with 123 participants, including students, educators, and active law enforcement officers, validated the system’s efficacy. Quantitative results demonstrate high user acceptance for Response Speed (4.7/5), Ease of Use (4.4/5), and Accuracy (4.3/5). {Qualitative feedback indicates that the agentic architecture is perceived as highly effective in guiding learners through personalized paths, demonstrating the feasibility and usability of agentic AI for specialized professional education.}

%Quantitative results demonstrate high user acceptance for Response Speed (4.7/5), Ease of Use (4.4/5), and Accuracy (4.3/5). Qualitative feedback confirms that the agentic architecture successfully fosters deep comprehension through personalized learning paths, offering a scalable solution for specialized professional education.
\end{abstract}


\section{Introduction}
The integration of Artificial Intelligence (AI) into educational curriculums has shifted from a novelty to a necessity. In the specialized field of Cybersecurity for Criminal Justice, the demand for accessible, accurate, and context-aware operational knowledge is critical. While Large Language Models (LLMs) like ChatGPT have democratized access to information, their unguided application in education presents significant challenges. Recent longitudinal studies indicate that uncontrolled access to generic LLMs can detrimentally affect students' long-term learning outcomes and exam performance \cite{ohm2024assessing}. Generic models often function as passive information retrieval tools, suffering from statelessness—a lack of continuity that prevents them from guiding learners through complex, longitudinal cognitive trajectories. Furthermore, the specific needs of criminal justice professionals, ranging from students focusing on conceptual legal theories to police officers requiring immediate operational procedures, cannot be adequately met by a ``one-size-fits-all'' reactive model.

The core challenge, therefore, is to leverage AI capabilities to maximize pedagogical effectiveness. Achieving this goal requires a system that moves beyond simple reactivity. It demands a robust architecture that ensures domain accuracy to prevent hallucinations, while simultaneously implementing sophisticated strategies like scaffolding and reflective learning—essential components for fostering deep comprehension \cite{zhang2025prompttutor, li2025can}.

To address these gaps, we propose the CyberJustice Tutor, a novel educational dialogue system designed specifically for the intersection of cybersecurity and criminal justice. Unlike standard chatbots, our system is built upon an Agentic AI framework. Central to this architecture is an autonomous pedagogical agent that employs a ``Think–Plan–Act'' cognitive cycle. This allows the system to analyze user intent, autonomously decompose high-level learning goals into structured plans, and execute dynamic pedagogical strategies. We introduce a Pedagogical Scaffolding Layer grounded in Vygotsky's Zone of Proximal Development (ZPD), which adapts the instructional support level based on the user's real-time progress. Furthermore, to mitigate the risk of misinformation in high-stakes legal contexts, we implement an Adaptive Retrieval Augmented Generation (RAG) core, grounding the agent's reasoning in verified curriculum materials.

This paper presents the design and evaluation of the CyberJustice Tutor. Our contributions are as follows:
\begin{enumerate}
    \item \textbf{Agentic Cognitive Architecture:} We introduce a novel framework that utilizes a cognitive cycle (Think-Plan-Act) to enable autonomous goal decomposition and longitudinal planning, transcending the limitations of reactive LLMs.
    \item \textbf{Pedagogical Alignment with ZPD:} We demonstrate the implementation of dynamic scaffolding strategies within an AI agent, ensuring interactions are cognitively aligned with the learner's developmental zone.
    \item \textbf{Empirical Evaluation:} {We report on a comprehensive user study with novice learners and experts, providing empirical evidence of the system's usability, acceptance, and perceived pedagogical utility, thereby validating the potential of agentic architectures in specialized educational contexts.}
    
    %We report on a pilot user study with novice learners, providing empirical evidence that the agentic architecture achieves high usability and perceived accuracy while effectively mediating the learning process through personalized plans.
\end{enumerate}

\section{Related Work}

The architecture of the CyberJustice Tutor is situated at the convergence of generative artificial intelligence in education, the paradigm of agentic systems, and established cognitive learning theories.

\subsection{Generative AI and Flipped Learning in Education}
The integration of Large Language Models (LLMs) into educational curricula represents a paradigm shift from static content delivery to dynamic interaction. Recent empirical research by Brito et al. \cite{brito2025enhancing} demonstrated that integrating AI-focused modules into cybersecurity curricula could increase student knowledge retention by 30\%. This supports the "Flipped Classroom" model, where direct instruction shifts to pre-class engagement. While Akçayır and Akçayır \cite{akcayir2018flipped} noted that flipped learning enhances engagement, it historically suffered from a lack of immediate, personalized feedback. 
To address this, the field has moved towards AI-driven tutoring. However, unguided reliance on generic models presents significant risks. Ohm et al. \cite{ohm2024assessing} found that uncontrolled ChatGPT usage correlated with lower exam scores due to surface-level processing. Furthermore, as highlighted by Bengio et al. \cite{bengio2023ai} and Kasneci et al. \cite{kasneci2023chatgpt}, generic models often hallucinate or lack domain-specific grounding. To bridge this gap, effective systems must emulate the "Socratic" method utilized by expert human tutors \cite{chi2001socratic}, prompting deep reasoning rather than merely providing answers.

\subsection{The Shift from AI Agents to Agentic AI}
Our system architecture is informed by the rapidly evolving taxonomy of autonomous systems. Recent surveys by Wang et al. \cite{wang2023survey} and Xi et al. \cite{xi2023rise} distinguish between traditional AI Agents and Agentic AI, which exhibits system-level intelligence through autonomous goal decomposition and persistent memory. Early iterations, such as AutoGPT \cite{yang2023autogpt} and BabyAGI \cite{nakajima2023babyagi}, demonstrated autonomous task execution but struggled with stability. 
Current frameworks emphasize multi-agent orchestration. Wu et al. \cite{wu2023autogen} introduced AutoGen for conversational programming, while Li et al. \cite{li2023camel} proposed the CAMEL framework to reduce hallucinations through role-playing. Similarly, Qian et al. \cite{qian2023chatdev} developed ChatDev to simulate standardized operating procedures (SOPs) through decentralized collaboration. Wang et al. \cite{wang2023voyager} further pushed the boundary with Voyager, demonstrating lifelong learning in embodied agents. Our CyberJustice Tutor adopts this \textit{Agentic AI} paradigm \cite{sapkota2025ai}, employing a unified cognitive workflow to handle the complex reasoning required in criminal justice scenarios.

\subsection{Cognitive Architectures: Reasoning and RAG}
To ensure reliability in high-stakes domains, robust cognitive architectures are essential. The ReAct framework by Yao et al. \cite{yao2023react} synergizes reasoning and acting, a principle central to our "Think–Plan–Act" cycle. This is complemented by Chain-of-Thought (CoT) prompting \cite{wei2022chain}, which significantly improves reasoning performance. Kojima et al. \cite{kojima2022large} further demonstrated that LLMs are "zero-shot reasoners" when properly prompted.
However, to mitigate the critical bottleneck of hallucination \cite{ji2023survey}, RAG conceptualized by Lewis et al. \cite{lewis2020retrieval}, has become standard. Simoni et al. \cite{simoni2025morse} utilized RAG in the MoRSE chatbot to anchor cybersecurity answers in verified databases. Additionally, Schick et al. \cite{schick2023toolformer} showed that LLMs can autonomous invoke external tools to verify facts. By combining CoT reasoning with RAG, our system ensures pedagogical guidance is legally accurate.

\subsection{Pedagogical Scaffolding and Persona Design}
Finally, the efficacy of an educational agent depends on its pedagogical grounding. Our approach is rooted in scaffolding, originally defined by Wood, Bruner, and Ross \cite{wood1976role}. Zhang et al. \cite{zhang2025prompttutor} validated this in the AI context with "PromptTutor," proving that scaffolding strategies improve learning outcomes compared to direct answers. This aligns with Wiggins and McTighe's \cite{wiggins2005understanding} backward design framework, mirrored by our agent's planning module.
This support is enhanced by effective persona design. Park et al. \cite{park2023generative} demonstrated that agents with specific personas can simulate believable social behaviors. In education, studies by Lieb and Goel \cite{lieb2024student} and Lo et al. \cite{lo2025noel} indicate that consistent personas increase engagement. Furthermore, motivation is critical; as outlined in the MSLQ by Pintrich et al. \cite{pintrich1991manual}, intrinsic value is a key driver of learning. Our system integrates these findings by adopting a Senior Cybercrime Analyst persona that socially mediates the learning process.

\section{Methodology}

The design of the CyberJustice Tutor represents a convergence of advanced computational intelligence and socio-cultural educational theory. Our methodology moves beyond the paradigm of reactive chatbots to establish an Agentic AI framework capable of autonomous planning and dynamic pedagogical intervention. This section delineates the theoretical foundations of Agentic AI and Vygotskian Scaffolding, followed by their technical realization in our unified system architecture.

\subsection{Agentic AI Framework and Cognitive Planning}

The architectural foundation of the CyberJustice Tutor is predicated on the principles of Agentic AI, a paradigm defined by Sapkota et al. \cite{sapkota2025ai} as systems capable of autonomous goal decomposition, dynamic orchestration, and persistent memory across complex workflows. Unlike traditional AI agents that operate as modular, task-specific entities executing isolated operations via tool interfaces, our system leverages Agentic AI to exhibit system-level intelligence. This distinction is critical for educational contexts, where effective tutoring requires not merely retrieving static information but constructing a coherent, longitudinal learning trajectory that adapts to the learner's evolving needs.

To operationalize this capability, the system employs a ``Think–Plan–Act'' cognitive cycle that shifts the locus of control from a linear prompt-response mechanism to an emergent, self-correcting loop. When a novice learner poses a high-level, unstructured query—such as asking how to investigate a complex cybercrime scenario—the system does not simply retrieve a generic list of steps. Instead, it utilizes its reasoning depth to decompose this abstract goal into a structured pedagogical sequence, defining necessary prerequisites before introducing advanced forensic concepts. This capacity for multi-step reasoning allows the system to handle the ambiguity inherent in student inquiries, bridging the gap between novice intent and expert knowledge structure.

Furthermore, the Agentic AI framework ensures temporal continuity and proactivity throughout the educational dialogue. By maintaining a persistent state of the learner's progress, the system moves beyond passive reactivity to actively manage the conversation flow. The architecture proactively recalls prior misconceptions and adjusts the current pedagogical strategy accordingly, creating a cohesive educational narrative rather than a disjointed series of independent interactions. This sustained context awareness is essential for mimicking the guidance of a human mentor, ensuring that the AI’s interventions are consistently aligned with the learner’s long-term educational objectives.

{We distinguish this framework from general-purpose automation by specializing the agent's capabilities. Rather than prioritizing open-ended external actions (e.g., browsing the live web or executing software commands), instead, we position our system as exhibiting pedagogical agency. This refers to the autonomy to alter instructional strategies without explicit user prompting. The key novelty lies in longitudinal pedagogical planning—the system’s capacity for persistent goal tracking, autonomous plan revision, and adaptive scaffolding across multi-turn learning trajectories. In this context, the environment the agent manipulates is not a physical interface, but the learner's cognitive state.}

\subsection{Pedagogical Alignment: ZPD and Instructional Scaffolding}

To ensure that the autonomous capabilities of the Agentic AI translate into effective learning outcomes, the system’s interaction model is grounded in Vygotsky’s ZPD. As discussed by Shabani et al. \cite{shabani2010vygotsky}, the ZPD represents the critical distance between a learner’s actual developmental level, determined by independent problem solving, and their potential development under expert guidance. Within the context of the CyberJustice Tutor, the Agentic AI assumes the role of the ``More Knowledgeable Other,'' mediating this cognitive transition. Rather than providing static information dumps which often lead to cognitive overload, or withholding too much information which results in frustration, the system aims to operate precisely within this zone, offering assistance that is contingent on the learner's real-time performance.

This theoretical framework is operationalized through a dynamic strategy of instructional scaffolding, where the system controls elements of the task initially beyond the learner's capacity while allowing them to concentrate on manageable components. The interaction follows a graduated progression of support, effectively fading assistance as the learner gains competence.(see Table \ref{tab:scaffolding}) For novice users encountering new concepts, the system initiates a high-support model characterized by direct instruction and explicit modeling of reasoning processes. As the learner demonstrates comprehension, the agent transitions to a collaborative mode, offering strategic hints and partial templates that encourage the user to bridge logical gaps independently. Finally, to verify the internalization of concepts, the system withdraws support to pose open-ended challenges, requiring the learner to synthesize information and apply it autonomously to complex scenarios.

The integration of this pedagogical scaffolding with the underlying Agentic AI architecture enables a highly responsive educational experience. The system’s planning module continuously monitors the user’s inputs to diagnose their current standing within the ZPD. If a learner struggles with a specific concept, the agent detects the breakdown and dynamically reinstates scaffolding—breaking the problem down further or providing an analogy—before attempting to move forward again. This adaptive calibration ensures that the instructional pace is neither too trivial nor too overwhelming, fostering a state of flow that is essential for mastering the intricate legal and technical nuances of cybersecurity in criminal justice.

\begin{table}[htbp]
\caption{Adaptive Scaffolding Stages within the ZPD}
\begin{center}
\begin{tabular}{p{2.2cm} p{1.2cm} p{4.2cm}}
\toprule
\textbf{Support Level} & \textbf{Phase} & \textbf{Agentic Pedagogical Strategy} \\
\midrule
\textbf{High Support} \newline (Modeling) & \textit{``I Do''} & \textbf{Direct Instruction:} The agent explicitly defines concepts, models the reasoning process, and establishes the foundational knowledge base to reduce cognitive load. \\
\midrule
\textbf{Guided Support} \newline (Co-Construction) & \textit{``We Do''} & \textbf{Collaborative Inquiry:} The agent offers strategic hints, partial templates, and guiding questions, prompting the learner to bridge logical gaps with assistance. \\
\midrule
\textbf{Low Support} \newline (Autonomy) & \textit{``You Do''} & \textbf{Independent Application:} The agent withdraws scaffolding to pose open-ended scenarios and complex challenges, verifying internalization and skill transfer. \\
\bottomrule
\end{tabular}
\label{tab:scaffolding}
\end{center}
\end{table}

\subsection{System Architecture and Implementation}
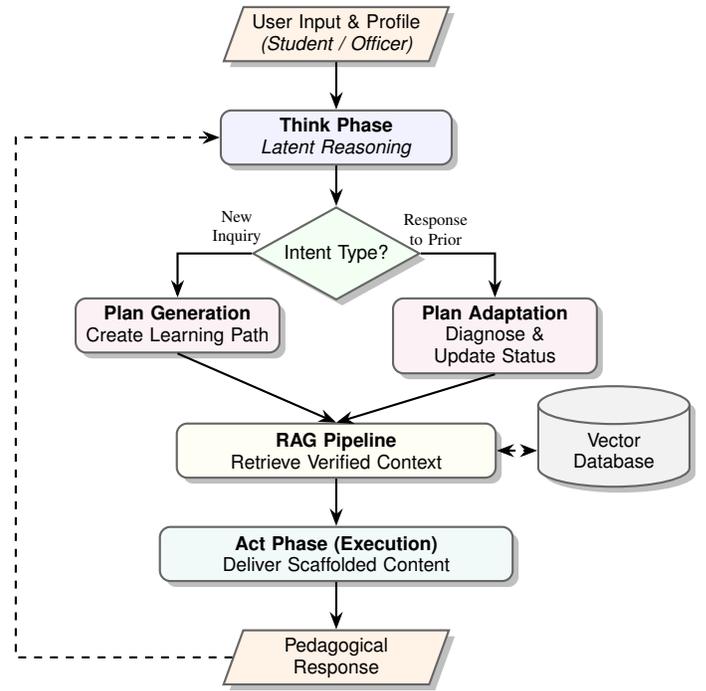
\begin{figure}[htbp]
\centering

\begin{tikzpicture}[
    scale=0.9, transform shape, % 整体缩放90%
    node distance=0.7cm, % 全局节点间距减小
    auto,
    >=Stealth,
    font=\footnotesize\sffamily, % 字体改小
    % 样式定义：减小了 text width, inner sep, minimum height
    process/.style={rectangle, draw=black!60, fill=blue!5, thick, text width=3.2cm, align=center, rounded corners, minimum height=0.8cm, inner sep=3pt, drop shadow},
    decision/.style={diamond, draw=black!60, fill=green!5, thick, text width=1.8cm, align=center, aspect=1.8, inner sep=1pt, drop shadow},
    io/.style={trapezium, trapezium left angle=70, trapezium right angle=110, draw=black!60, fill=orange!10, thick, text width=2.5cm, align=center, minimum height=0.8cm, inner sep=3pt, drop shadow},
    db/.style={cylinder, shape border rotate=90, draw=black!60, fill=gray!10, thick, aspect=0.25, text width=2cm, align=center, minimum height=1.2cm, drop shadow},
    group/.style={rectangle, draw=black!30, dashed, inner sep=5pt, rounded corners, fill=white}
]

    % --- Nodes ---

    % 1. Input
    \node (input) [io] {User Input \& Profile\\ \textit{(Student / Officer)}};

    % 2. THINK Phase
    \node (think) [process, below=of input] {\textbf{Think Phase}\\ \textit{Latent Reasoning}};

    % 3. PLAN Phase
    \node (decide) [decision, below=0.6cm of think] {Intent Type?};
    
    % Left Branch
    \node (decompose) [process, below left=0.3cm and 0.2cm of decide, fill=purple!5, text width=2.8cm] {\textbf{Plan Generation}\\ Create Learning Path};
    
    % Right Branch
    \node (assess) [process, below right=0.3cm and 0.2cm of decide, fill=purple!5, text width=2.8cm] {\textbf{Plan Adaptation}\\ Diagnose \& Update Status};

    % 4. RAG Core
    \node (rag) [process, below=1.8cm of decide, text width=4.5cm, fill=yellow!5] {\textbf{RAG Pipeline}\\ Retrieve Verified Context};
    
    % Database 
    \node (database) [db, right=0.6cm of rag] {Vector Database};

    % 5. ACT Phase
    \node (act) [process, below=of rag, text width=5cm, fill=teal!5] {\textbf{Act Phase (Execution)}\\ Deliver Scaffolded Content};

    % 6. Output
    \node (output) [io, below=of act] {Pedagogical Response};

    % --- Connections ---
    \draw[->, thick] (input) -- (think);
    \draw[->, thick] (think) -- (decide);
    
    % Decision paths: 
    \draw[->, thick] (decide.west) -| node[above, pos=0.1, align=center, font=\scriptsize] {New\\Inquiry} (decompose.north);
    \draw[->, thick] (decide.east) -| node[above, pos=0.1, align=center, font=\scriptsize] {Response\\to Prior} (assess.north);
    
    % Converge to RAG
    \draw[->, thick] (decompose.south) -- (rag.north);
    \draw[->, thick] (assess.south) -- (rag.north);
    
    % RAG to DB
    \draw[<->, dashed, thick] (rag) -- (database);
    
    % RAG to ACT
    \draw[->, thick] (rag) -- (act);
    
    % ACT to Output
    \draw[->, thick] (act) -- (output);
    
    % Feedback Loop: 
    \draw[->, dashed, thick] (output.west) -- ++(-3.2,0) |- node[near start, above] {} (think.west);

\end{tikzpicture}
\caption{The architectural workflow of the CyberJustice Tutor. It classifies user intent (Latent
Reasoning), generates learning paths or analyzes gaps (Planning), and delivers verified content aligned with the user’s ZPD (Execution)}
\label{fig:architecture}
\end{figure}

The CyberJustice Tutor is engineered as a decoupled web application orchestrated via the LangChain framework, utilizing OpenAI's GPT-4o as the core reasoning engine for its superior logical deduction capabilities.As illustrated in Figure \ref{fig:architecture}, upon receiving input, the agent initiates a latent reasoning phase to classify the learner's intent—distinguishing between a new inquiry and a response to a prior scaffold. If a significant knowledge gap is detected, the planning module autonomously generates a multi-step learning path, ensuring that the subsequent execution phase delivers content that is structurally aligned with the user's current Zone of Proximal Development.

To mitigate the hallucinations common in generative models and ensure domain accuracy, the agent’s reasoning is grounded in a RAG pipeline. The system maintains a specialized knowledge base comprising comprehensive curriculum materials, ranging from legal statutes to digital forensics procedures. By retrieving the top semantically relevant chunks prior to generation, the system effectively constrains the LLM's output to the verified knowledge base, minimizing hallucination risks in high-stakes contexts.

Finally, the interaction is mediated through a ``Senior Cybercrime Analyst'' persona, designed to foster professional immersion. To ensure the system evolves alongside its users, a Human-in-the-Loop feedback mechanism is embedded directly into the interface. Explicit user ratings and implicit interaction data are logged to a structured dataset, creating a feedback loop that allows for the iterative refinement of both the RAG retrieval relevance and the agent’s scaffolding strategies, thereby maintaining the system's efficacy as an adaptive educational tool.

To comprehensively illustrate how the Agentic AI, Pedagogical Scaffolding, and System Architecture converge in a real-world interaction, Figure \ref{fig:malware_case} presents a detailed operational walkthrough. This transcript highlights the agent's internal cognitive cycle, from decomposing the abstract topic of `Malware Defense' into a structured learning plan, to applying specific scaffolding strategies based on the user's responses.

\begin{figure*}[htbp]
\centering
\begin{tcolorbox}[colback=white, colframe=black!75, title=\textbf{Case Study: Agentic Planning for Malware Defense}]
\small
\textbf{Context:} User is identified as a \textit{CJ Student} (Novice). The goal is to learn defensive strategies.\\
\rule{\linewidth}{0.4pt}

\textbf{User:} ``What should I do to defend against malware?''

\vspace{0.2cm}
\textit{\textbf{[Agent Internal State]}}
\begin{itemize}
    \item \textbf{Intent:} New Topic (Malware Defense) $\rightarrow$ \textit{Plan Generation}.
    \item \textbf{Plan Created (5 Steps):} 
    1. Definitions \& Vectors; 2. Baseline Hygiene; 3. Anti-malware Tools; 4. Layered Defenses; 5. Incident Response.
    \item \textbf{Current Step:} Step 1 (Definitions).
    \item \textbf{Scaffolding Strategy:} \textbf{High Support (``I Do'')}. Establish foundational concepts before tactics.
\end{itemize}

\textbf{Agent:} ``Great question! To defend effectively, we first need to understand the threat.
\newline
\textbf{Step 1: Fundamentals.} Malware includes various categories like \textit{Viruses} (attach to files), \textit{Worms} (self-replicate), and \textit{Ransomware} (encrypt data).
\newline
\textit{Check for Understanding:} Can you name at least three types you've heard of and describe how they differ? Feel free to guess—we will clarify misconceptions together.''

\vspace{0.2cm}
\rule{\linewidth}{0.1pt} % Separator
\vspace{0.2cm}

\textbf{User:} \\
``1. \textbf{Virus:} Attaches to a file and needs me to open it.
2. \textbf{Ransomware:} Locks my files until I pay money.
3. \textbf{Trojan:} Disguises itself as fake software so I install it.''

\vspace{0.2cm}
\textit{\textbf{[Agent Internal State]}}
\begin{itemize}
    \item \textbf{Intent:} Response to Scaffold $\rightarrow$ \textit{Plan Adaptation}.
    \item \textbf{Evaluation:} Definitions are accurate. Mastery of Step 1 core concepts confirmed.
    \item \textbf{Next Action:} Complete Step 1 (Infection Vectors) before moving to Step 2 (Hygiene).
    \item \textbf{Scaffolding Strategy:} \textbf{Guided Support (``We Do'')}. Validate success and collaboratively fill the remaining knowledge gaps (Worms/Rootkits).
\end{itemize}

\textbf{Agent:} ``Spot-on! You have correctly identified the execution mechanism of a Virus versus the deception of a Trojan.
\newline
Before we move to \textbf{Step 2 (Hygiene Practices)}, we need to cover the missing pieces: \textit{Worms} (network spread) and \textit{Rootkits} (hidden access). Let's look at how these typically infect a system...''

\end{tcolorbox}
\caption{An operational instantiation of the Agentic Tutor Framework. This example illustrates how the system's \textit{Planning Module} translates a high-level user intent (``Defend against malware'') into a structured 5-step curriculum, while the \textit{Execution Phase} applies high-support scaffolding logic as defined in Table \ref{tab:scaffolding}.}
\label{fig:malware_case}
\end{figure*}

\section{Experiment Design and Evaluation}
To validate the architectural integrity and pedagogical efficacy of the CyberJustice Tutor, we conducted a comprehensive user study focused on three critical dimensions: the usability of the agentic interface, the factual accuracy of the RAG-generated content, and the effectiveness of the dynamic scaffolding strategies.

\subsection{Participants and Study Protocol}
The study cohort comprised 123 participants recruited from Criminal Justice and Technology programs, representing a diverse cross-section of the target user base. The demographic distribution included undergraduate CJ students (approx. 75\%), active law enforcement officers, legal professionals, and educators. This diversity is crucial; while students represent the "Novice" ZPD, professionals and educators serve as domain experts to validate the accuracy of the content. Notably, 65\% of participants reported only occasional familiarity with AI tools, ensuring the study measured the system's intuitive design rather than the users' prior AI proficiency.

To evaluate the system in a realistic educational context, we adopted an open-ended study protocol. Rather than enforcing a rigid experimental script or fixed time constraints, participants were granted access to the CyberJustice Tutor and encouraged to use it as a supplementary learning tool for their coursework or professional development. Consequently, engagement sessions were self-directed, with usage durations varying naturally based on individual user needs. Participants were invited to freely explore the system's core capabilities, such as querying foundational concepts or engaging in investigative scenarios, allowing us to assess how the agent adapts to diverse, spontaneous user intents.

\section{Results}
Data was collected via a post-interaction survey utilizing both Likert-scale ratings (1--5) and open-ended feedback mechanisms to capture both the performance metrics and the user experience of the agentic interaction.

\subsection{Quantitative Performance}
The quantitative results, summarized in Table \ref{tab:results}, indicate a strong and consistent user acceptance of the backend agentic intelligence across a large sample size (\textit{N=123}). The inclusion of educators and practitioners in the expanded cohort provides significant validation of the system's professional utility.

\begin{table}[htbp]
\caption{User Evaluation Ratings (Scale 1--5, N=123)}
\begin{center}
\begin{tabular}{lccc}
\toprule
\textbf{Metric} & \textbf{Mean} & \textbf{Std. Dev} & \textbf{Interpretation} \\
\midrule
Response Speed & 4.7 & 0.5 & Low latency routing \\
Ease of Use & 4.4 & 0.7 & High intuitive adoption \\
Accuracy & 4.3 & 0.6 & RAG effectively verifies facts \\
Relevance & 4.2 & 0.7 & Context-aware outputs \\
Practicality & 4.1 & 0.8 & Actionable guidance \\
Visual Appeal & 3.5 & 1.1 & Functional but needs polish \\
\bottomrule
\end{tabular}
\label{tab:results}
\end{center}
\end{table}

The system achieved exceptionally high marks for \textbf{Response Speed (4.7)} and \textbf{Ease of Use (4.4)}. This performance data validates the efficiency of the Agent architecture, confirming that the ``Think–Plan–Act'' cognitive cycle operates without introducing perceptible latency. Notably, the \textbf{Accuracy (4.3)} rating reinforces the effectiveness of the RAG pipeline in mitigating hallucinations. 

{Regarding Visual Appeal, the moderate score ($3.5 \pm 1.1$) reflects the system's status as a functional research prototype rather than a commercial product. The research priority was explicitly placed on engineering the backend agentic logic and scaffolding mechanisms. The higher standard deviation indicates a divergence in user expectations; however, the high Ease of Use score suggests that while the interface lacked aesthetic polish, it successfully facilitated the intended pedagogical interactions without friction.}

\subsection{Qualitative Validation of Methodology}
The open-ended feedback provided empirical support for the specific pedagogical design choices embedded in the system's cognitive architecture.

Primary validation for the \textbf{Scaffolding Strategy} emerged from user comments regarding the instructional pacing. One educator explicitly noted: \textit{``The tutor provided clear and organized explanations, and more importantly, asked me questions in `next steps for you' that inspired me to think further and guided me to a learning flow.''} This directly confirms that the agent successfully operated within the users' Zone of Proximal Development, moving beyond passive answering to active cognitive guidance. Another participant highlighted the system's ability to \textit{``translate technical cybersecurity concepts into clear, criminal justice-focused language,''} validating the effectiveness of the Persona design in bridging domain gaps.

Furthermore, the \textbf{Agentic Planning Capabilities} were highlighted as a key differentiator. Users described the generated Learning Plans as providing good structure and helping them learn systematically. A novice user reported that the system \textit{``helped me understand network intrusion basics by breaking down a complex log-analysis scenario into simple, step-by-step explanations.''} This feedback confirms that the Planning Module successfully decomposed high-level abstract goals into actionable educational sequences, a core capability of Agentic AI that static chatbots often lack.

%\subsection{Discussion}
%The findings present a compelling counter-narrative to Ohm
%et al. [1], who cautioned that uncontrolled access to generic
%LLMs could degrade learning outcomes. Our results suggest
%that when AI is constrained by an Agentic framework and
%grounded by RAG, it shifts from a passive information retrieval
%tool to an active, pedagogical partner. The users’ preference for
%the system’s “High Interactivity” and “Practical Focus” (Mean
%4.1) indicates that the system successfully fostered social inter
%subjectivity. By combining the precision of a retrieval engine
%with the adaptive empathy of a human mentor, the CyberJus
%tice Tutor demonstrates that domain-specific customization is
%essential for professional education in high-stakes fields.

\section{Conclusion}
{This study investigated the feasibility of integrating an Agentic AI framework into cybersecurity education for criminal justice. By combining a ``Think–Plan–Act'' cognitive cycle with Vygotskian scaffolding and RAG, we addressed the limitations of generic LLMs regarding continuity and hallucinations. Our findings from a 123-person user study demonstrate that the CyberJustice Tutor achieves high marks for usability and perceived accuracy. While we cannot yet claim quantified learning gains without longitudinal testing, the strong user acceptance and qualitative feedback suggest that the system effectively engages learners and adapts to their cognitive needs. This work establishes a foundational architecture for agentic pedagogical systems, proving that autonomous planning and dynamic scaffolding are viable and valuable additions to AI-driven education. Future work will focus on controlled trials to measure long-term knowledge retention.}

\section*{Acknowledgment}

We sincerely thank criminal justice students and professionals for their participation in this anonymous survey on the use of chatbots.

% \section*{References}

% Please number citations consecutively within brackets \cite{b1}. The 
% sentence punctuation follows the bracket \cite{b2}. Refer simply to the reference 
% number, as in \cite{b3}---do not use ``Ref. \cite{b3}'' or ``reference \cite{b3}'' except at 
% the beginning of a sentence: ``Reference \cite{b3} was the first $\ldots$''

% Number footnotes separately in superscripts. Place the actual footnote at 
% the bottom of the column in which it was cited. Do not put footnotes in the 
% abstract or reference list. Use letters for table footnotes.

% Unless there are six authors or more give all authors' names; do not use 
% ``et al.''. Papers that have not been published, even if they have been 
% submitted for publication, should be cited as ``unpublished'' \cite{b4}. Papers 
% that have been accepted for publication should be cited as ``in press'' \cite{b5}. 
% Capitalize only the first word in a paper title, except for proper nouns and 
% element symbols.

% For papers published in translation journals, please give the English 
% citation first, followed by the original foreign-language citation \cite{b6}.

\bibliographystyle{IEEEtran}
\bibliography{ref}

\end{document}